\documentclass[lineno]{jfm}

\usepackage{graphicx}
\usepackage{float}
\usepackage{bm}
\usepackage{newtxtext}
\usepackage{newtxmath}
\usepackage{natbib}
\usepackage{hyperref}
\usepackage[export]{adjustbox}
\hypersetup{
    colorlinks = true,
    urlcolor   = blue,
    citecolor  = black,
}

\newcommand{\RomanNumeralCaps}[1]

\title{Enhanced magnetostrophic waves with magnetic field orthogonal 
to the rotation  axis}

\author{Raviraj Narayan Shinde \aff{1}
  \corresp{\email{ravirajs@iisc.ac.in}},
Ghanesh Narasimhan\aff{2}
 }

\affiliation{\aff{1}Centre for Earth Sciences, Indian Institute of Science, Bangalore 560012, India
\aff{2} Department of Mechanical Engineering \& St. Anthony Falls Lab, University of Minnesota, Minneapolis, 55414, USA}

\begin{document}
\maketitle

\begin{abstract}

We consider an isolated Gaussian velocity vortex perturbation in an otherwise quiescent, electrically conducting, and rotating fluid permeated by a uniform magnetic field $\bm{B}$. Studies suggest a presence of strong azimuthal wave motions on the timescale of centuries within the Earth's liquid outer core at higher latitudes. To understand these long-period oscillations, we focus on magnetostrophic waves, a slow component of magnetic-Coriolis waves with $\bf{B}$ orthogonally aligned with the rotation vector $\bf{\Omega}$, which replicate the field lines in the azimuthal direction. We present an analytical solution to the magnetic-Coriolis wave equation in Cartesian coordinates. Later, with numerical solution, we validate our analytical estimates and show that magnetostrophic waves travel relatively faster along the magnetic field when $\bm{B} \perp \bf{\Omega}$ compared with the case when both are aligned. The study confirms that with a magnetic field $\bm{B}$ orthogonally aligned to the rotation vector $\bm{\Omega}$, wave vectors satisfying the condition $\bm{\Omega}\cdot\bm{k} \approx 0$, travel with Alfvén velocities along the magnetic field lines as a component of inertial-Alfvén waves. The timescales on which Alfvén waves travel are relatively short, and it is also less likely that inertial-Alfvén waves will be sustained inside the core at higher latitudes \citep{Davidson2017}. This study shows that, excluding the inertial-Alfvén waves contribution ($k_z\neq 0$), there exists intensified magnetostrophic wave propagation when $\bm{B} \perp \bf{\Omega}$, which can explain the strong periodic oscillations on the time scales of centuries along the azimuthal field at higher latitudes.
Results show persistence of the magnetostrophic waves despite the lower Lehnert number $Le$, suggesting the plausible existence of a low-intensity azimuthal magnetic field in the Earth's core.

\end{abstract}




\section{Introduction}\label{sec:1}
Study of magnetohydrodynamic (MHD) waves has proven helpful in inferring magnitudes of inaccessible magnetic fields inside the Earth's core \citep{Gillet2010}. Particularly, the slow magnetostrophic waves can help explain non-axisymmetric oscillations that travel in azimuthal directions with a periodicity of a few hundred years \citep{Hori2015}. As these waves travel in an azimuthal direction (East-West) along the magnetic field perpendicular to the rotation axis, it is necessary to investigate rotating turbulence in a magnetic field $\bm{B}$ orthogonal to the rotation axis to make such inferences. 

\cite{Finlay2010} and  \cite{Hori2022} in their reviews mentioned that slow magnetostrophic waves might have different properties depending on the orientation of the magnetic field relative to the rotation axis. 
\citet{Davidson2017} studied the magnetic-Coriolis (MC) waves with $\bm{B} \perp \bm{\Omega}$ and have shown that there exist at least four types of waves, namely weakly modified inertial waves, slow magnetostrophic waves, inertial-Alfv\'{e}n waves and intermediate magnetic-Coriolis waves. Out of these, inertial-Alfv\'{e}n waves and intermediate magnetic-Coriolis waves are specifically noticed when the magnetic field $\bm{B}$ is orthogonal to rotation $\bm{\Omega}$. Their study highlights the propagation of waves along the rotation axis and the possibility of forming columnar structures aligned with it within the Earth's core. However, the study does not demonstrate how these newly discovered waves impact travel along the magnetic field lines within the core. 

In this study, comparing the cases $\bm{B}\perp \bm{\Omega}$ and $\bm{B}\parallel \bm{\Omega}$, we show the variations in the velocity of travel along the magnetic field lines due to magnetostrophic waves. We also discuss the importance of these waves for understanding periodic variations in the azimuthal magnetic field inside the Earth's core at higher latitudes. Based on the results obtained, we also comment on the intensity of the internal azimuthal magnetic field.

In the following Section \ref{sec:2}, we will analytically solve an initial value problem to see the slow magnetostrophic wave's evolution in the wave-dominated phase. Following that, in Section \ref{sec:3} we discuss corresponding results based on the numerical simulation. Lastly, in Section \ref{sec:4}, we will discuss the implications of the results for the Earth's core.
\section{Analytical solution to slow magnetostrophic wave's velocity}\label{sec:2}
\subsection{Dispersion relation for the magnetic-Coriolis waves }
Consider an electrically conducting incompressible fluid undergoing uniform rotation $|\bm{\Omega}| \bm{\hat{e}_z}$ in the presence of an applied uniform magnetic field $\bm{B}$. Suppose this flow is characterized by a low magnetic Reynolds number $R_m \ll 1$ ($R_m$ is ratio of the magnetic field’s advection to the magnetic field’s diffusion), low Rossby number $R_o \ll 1$ ($R_o$ is ratio of magnitudes of inertia force to Coriolis force), and a large magnetic interaction parameter $N \gg 1$ ($N$ is ratio of magnitudes of the Lorentz force to the Inertia force). In the context of Earth's outer core, the magnetic Prandtl number ($\nu/\eta$) is of the order of $10^{-6}$, allowing us to neglect the effects of kinematic viscosity ($\nu$) compared to magnetic diffusivity ($\eta$). Under these assumptions, we can express the linearised governing momentum equation and the magnetic induction equation in a simplified form as follows,
\begin{equation}\label{eq:1.1}
    \frac{\partial \bm{u}}{\partial t}= -\frac{\nabla p'}{\rho}  -2  \bm{\Omega} \times   \bm u+ \frac{(\bm{B}\cdot \nabla) \bm{b}}{\rho\mu}.
\end{equation}
\begin{equation}\label{eq:1.2}
\frac{\partial\bm{b}}{\partial t} = (\bm{B}\cdot \nabla)  \bm{u}  + \eta \nabla^2 \bm{b}.   
\end{equation}
On the right-hand side of \eqref{eq:1.1}, the first term contains all gradient terms, including the magnetic pressure $\textbf{b}^2/2\mu$. The second and third terms correspond to the Coriolis and Lorentz forces per unit mass, respectively, with ${\rho}$ denoting the mass density of the fluid. Here, $\bm{b}$ is the induced magnetic field and $\eta = (\mu \sigma)^{-1}$ is the magnetic diffusivity where $\mu$ and ${\sigma}$ are the permeability and electrical conductivity of the fluid, respectively. Solving \eqref{eq:1.1} and \eqref{eq:1.2} gets us the magnetic-Coriolis wave equation with the effect of magnetic diffusivity, \citep{Lehnert1954}, \citep{Sreenivasan17}, 
\begin{equation}\label{eq:1.3}
  \left[\dfrac{\partial}{\partial t}\left(\dfrac{\partial}{\partial t} -\eta \nabla^2 \right)-\dfrac{(\bm{B} \cdot \nabla )^2 } {\rho\mu}\right]^2 (\nabla ^2 \bm{u} )   +4(\bm \Omega \cdot \nabla)^2\left(\dfrac{\partial}{\partial t}- \eta \nabla^2 \right)^2\bm{u} = 0.
\end{equation}
Applying Fourier transform to \eqref{eq:1.3}, we get
\begin{equation}\label{eq:1.4}
  \left[ \dfrac{\partial}{\partial t}  \left(\dfrac{\partial}{\partial t}+ \eta k^2 \right)+V_A^2k_{B}^2\right]^2  \hat{\bm{u}} +\frac{4{\Omega}^2{ k_z}^2}{k^2}  \left(\dfrac{\partial}{\partial t}+ \eta k^2 \right)^2 \hat{ \bm{u}} = 0. 
\end{equation}
Here, in a rectangular Cartesian coordinate system, $k_x, k_y$ and $k_z$ represent the components of the wave vector $\bm{k}$ along unit vectors $\bm{\hat{e}_x}$, $\bm{\hat{e}_y}$ and $\bm{\hat{e}_z}$ respectively. $k^2 = k_x^2+k_y^2+k_z^2$, $\bf{V_A} = \bm{B}/\sqrt{\rho \mu}$ is the Alfv\'{e}n wave velocity and $k_{B}$ is the magnitude of the wave vector in the direction of $\bm{B}$. In this study, we are going to considering two cases, a) magnetic field $|\bm{B}|\bm{\hat{e}_z}$  aligned with rotation $\bm{\Omega}$ referred throughout as parallel configuration and, b) magnetic field $|\bm{B}|\bm{\hat{e}_x}$ orthogonal to rotation $\bm{\Omega}$ referred as perpendicular configuration.   Substituting a solution  $\hat{\bm{u}} = e^{i\varpi t}$ in \eqref{eq:1.4}, we get two pairs of solutions for the temporal frequency $\varpi$ as follows,
\begin{equation}\label{eq:1.5}
    \varpi_{1,2} = \pm\dfrac {\omega_I}{2} + \dfrac{i \omega_d}{2} \pm \sqrt{\omega_A^2+\left(\dfrac {\omega_I}{2}\mp \dfrac{i  \omega_d}{2}\right)^2}.
\end{equation}
\begin{equation}\label{eq:1.6}
    \varpi_{3,4} = \pm\dfrac {\omega_I}{2} + \dfrac{i\omega_d}{2} \mp \sqrt{\omega_A^2+\left(\dfrac {\omega_I}{2}\mp \dfrac{i \omega_d}{2}\right)^2}.
\end{equation}
Where, $\omega_I = {2\Omega k_z}/{k}$ is the inertial wave frequency, $ \omega_A=V_Ak_{B}$ is the Alfv\'{e}n wave frequency and $\omega_d =\eta k^2 $ is the magnetic dissipation frequency. 
Lehnert number $Le = V_A/2\Omega\delta \ll 1$ ($Le$ is ratio of  Alfv\'{e}n to inertial frequencies, $\delta$ is the characteristic length scale of the flow) ensures that the first pair $\varpi_{1,2}$ corresponds to inertial wave weakly influenced by a magnetic field called as the weakly modified inertial wave. The second pair $\varpi_{3,4}$ represents a slowly evolving push and pull between Lorentz and Coriolis forces and is hence termed as a slow magnetostrophic wave. 
Also, we consider the magnetic Ekman number $E_{\eta} 
\ll Le\ll 1$
($E_\eta = \eta/2\Omega\delta^2$ represents the ratio of the strength of magnetic dissipation to the Coriolis force). This ensures the existence of an initial magnetic-Coriolis wave phase followed by magnetic dissipation. It is known that $\varpi_{1,2}$ corresponds to the components that predominantly evolve along the rotation vector $\bm{\Omega}$ and support the formation of columnar structures aligned with it \citep{Ranjan2014}. We will not focus on those in this study.  \cite{Davidson97}; \cite{Siso2004}; \cite{Tigrine2019} have investigated changes in the behaviour of magnetohydrodynamic waves in parallel and perpendicular configurations. But they did not comment on slow magnetostrophic waves and their evolution. In this study, we will compare the evolution of the slow magnetostrophic wave in parallel and perpendicular configurations to highlight the discrepancies.
\subsection{Slow magnetostrophic wave's approximate frequency}
In a regime defined by $E_{\eta} \ll Le \ll 1$, let us consider the mathematical expression for the slow magnetostrophic wave's frequency $\varpi_{3,4}$ from \eqref{eq:1.6} and expand it as shown below,
\begin{equation*}
    \varpi_{3,4} = \pm{\dfrac{\omega_I}{2}} +i \dfrac{ \omega_d}{2} \mp  \dfrac{\omega_I}{2} \sqrt{\left(\dfrac{2\omega_A}{\omega_I}\right)^2+1-\left( \dfrac{\omega_d}{\omega_I}\right)^2 \mp i\left( \dfrac{2\omega_d}{\omega_I}\right)} ,
\end{equation*}
\begin{equation*}
\hspace{5mm}
\dfrac{\omega_d}{\omega_I}\ll{\left(\dfrac{2\omega_A}{\omega_I}\right)\ll 1 } , \hspace{10mm} 
\end{equation*}
$\varpi_{3,4}$ approximately becomes,
\begin{equation}\label{eq:1.8}
    \varpi_{3,4} \approx \pm\dfrac {\omega_I}{2} +i \dfrac{ \omega_d}{2} \mp \dfrac {\omega_I}{2} \sqrt{1+\left(\dfrac{2\omega_A}{\omega_I}\right)^2 \mp i\left( \dfrac{2\omega_d}{\omega_I}\right)},
\end{equation}
simplifying the terms in the square root (see Appendix \ref{appA1}),
\begin{equation*}
  \varpi_{3,4} \approx  \mp\dfrac{ \omega_A^2}{\omega_I}+i\omega_d.
\end{equation*}
Substituting $\omega_A^{\parallel} = V_Ak_z$ for  the parallel configuration; $\bm{B}$ and $  \bm{\Omega}$ both directed along z-axis and $\omega_A^{\perp} = V_Ak_x$ for the perpendicular configuration; $\bm{B}$ is along x-axis, $ \varpi_{3,4}$ becomes,
\begin{subequations}\label{eq:1.8ab}
\begin{equation}
\begin{aligned}\label{eq:1.8a}
 \varpi_{3,4}^{\parallel} \approx  \mp \frac{V_A^2k_z k}{2\Omega }+i\omega_d,&\quad 
\end{aligned}
\end{equation}
\begin{equation}
\begin{aligned}\label{eq:1.8b}
\varpi_{3,4}^{\perp} \approx  \mp\frac{V_A^2k_{x}^2 k}{2\Omega k_z}+i\omega_d.   &\quad 
\end{aligned}
\end{equation}
\end{subequations}

The real parts of $\varpi_{3,4}$ have distinct mathematical expressions that represent the change in the frequency and phase velocity depending on how the magnetic field ($\bm{B}$) is oriented relative to the rotation vector ($\bm{\Omega}$). \cite{Finlay2010};  \cite{Hori2022} in their review suggested this fundamental change in the behaviour of the slow magnetostrophic waves depending on the orientation of $\bm{B}$ relative to $\bm{\Omega}$.  Interestingly, the imaginary parts of $\varpi_{3,4}$ that correspond to decay, characterised by the magnetic dissipation frequency ($\omega_d$), are the same in the two configurations. They could not comment on this due to the consideration of an ideal case without magnetic dissipation. Based on this, we expect slow magnetostrophic waves to start decaying on the same timescales irrespective of the relative orientation of the magnetic field ($\bm{B}$) and rotation vector ($\bm{\Omega}$). We will see that this gets validated by the numerical results in Section \ref{sec:3}.

Following \cite{Bardsley16}, in the perpendicular configuration for wave vectors satisfying the condition $\bm{\Omega}\cdot \bm{k} \approx$ 0, \eqref{eq:1.6} reduces to the inertial-Alfv\'{e}n wave's dispersion relation,
\begin{equation}\label{eq:1.7}
       \varpi \approx \pm \omega_A^{\perp} + \dfrac{i \omega_d}{2}.
\end{equation}
These waves travel along the magnetic field lines with the Alfv\'{e}n velocity ($V_A$).
\cite{Davidson2017} showed that, due to the non-uniform structure of Earth's magnetic field, inertial-Alfv\'{e}n waves can get confined to low latitudes. Investigations on the azimuthal magnetic field inside the core have proposed strong wave components along the azimuthal direction, especially at higher latitudes \citep{Hori2015}. Motivated by this, we primarily focus 
on the slow magnetostrophic component in the perpendicular configuration, keeping in mind that the inertial-Alfv\'{e}n waves will not contribute to azimuthal travel of wave disturbances at higher latitudes.

\subsection{Initial value problem: slow magnetostrophic wave velocity}

Considering $\varpi_{1,2}$ and $\varpi_{3,4}$ from \eqref{eq:1.5} and \eqref{eq:1.6} respectively, we can write the general solution for $\hat{\bm{u}}$ in \eqref{eq:1.4} (similarly for $\hat{\bm{b}}$),
\begin{subequations}
\begin{equation}\label{eq:2.1a}
  {\hat{\bm{u}}} = A_1 e^{{i\varpi_1 t}} +  B_1 e^{i\varpi_2 t}+ C_1 e^{i\varpi_3 t}+ D_1 e^{i\varpi_4 t}.
\end{equation}
\begin{equation}\label{eq:2.2b}
  {\hat{\bm{b}}} = A_2 e^{{i\varpi_1 t}} +  B_2 e^{i\varpi_2 t}+ C_2 e^{i\varpi_3 t}+ D_2 e^{i\varpi_4 t} .
\end{equation}
\end{subequations}
Now let's also consider an initial value problem in an infinite domain of rotating electrically conducting incompressible fluid with rotation vector $|\bm{\Omega}| \bm{\hat{e}_z}$  and applied magnetic field $\bm{B}$. The initial velocity perturbation is a two-dimensional axisymmetric velocity vortex in rectangular Cartesian coordinates,
\begin{equation}\label{eq:2.3}
   \bm{u}_0=  y e^{-\dfrac{(x^2+y^2+z^2)}{\delta^2}}\hat{\bm{e}}_x - x e^{-\dfrac{(x^2+y^2+z^2)}{\delta^2}}\hat{\bm{e}}_y.
\end{equation}
The initial induced magnetic field $\bm{b}_0$ and induced currents $\bm{j}_0$ are set equal to 0 (since it takes a finite time $t >0$ to develop), 
\begin{equation}\label{eq:2.4}
    \bm{b}_{0} = 0, \hspace{3mm}\bm{j}_0 =0. 
\end{equation}
Let us consider the equation \eqref{eq:2.1a} just for it's x component (except for the initial transition phase components along y and z direction evolve similarly) along with the  $\hat{\bm{u}}_x'$, $\hat{\bm{u}}_x''$, $\hat{\bm{u}}_x'''$ at $t= 0$ (where $'$ is time derivative).
\begin{subequations}\label{eq:2.4abcd}
\begin{equation}
    A_{1x} +B_{1x} +C_{1x}+D_{1x} =\left(\hat{u}_{x}\right)_{t=0}  = a_{1} .
\end{equation}
\begin{equation}
    A_{1x}\varpi_1 +B_{1x}\varpi_2+C_{1x}\varpi_3+D_{1x}\varpi_4 = -i\left(\dfrac{\partial \hat{u}_x }{\partial t}\right)_{t=0}  =  a_{2} .
\end{equation}

\begin{equation}
    A_{1x}\varpi_1^2 +B_{1x}\varpi_2^2+C_{1x}\varpi_3^2+D_{1x}\varpi_4^2 = -\left(\dfrac{\partial^2 \hat{u}_x }{\partial t^2}\right)_{t=0}  =  a_{3} .
\end{equation}

\begin{equation}
    A_{1x}\varpi_1^3 +B_{1x}\varpi_2^3+C_{1x}\varpi_3^3+D_{1x}\varpi_4^3 = i\left(\dfrac{\partial^3 \hat{u}_x }{\partial t^3}\right)_{t=0}   =  a_{4}.
\end{equation}
\end{subequations}
Using \eqref{eq:1.1} and \eqref{eq:1.2} we solve further and get,
\begin{subequations}\label{eq:2.5abcd}
\begin{equation}
    a_{1} = \hat{u}_{x0} ,  
\end{equation}
\begin{equation}
    a_{2} = i \omega_I ({k_y\hat{u}_{z0}-k_z\hat{u}_{y0}})/{k},
\end{equation}
\begin{equation}
    a_{3} = ( \omega_A^2 + \omega_I^2) \hat {u}_{x0},
\end{equation}
\begin{equation}
    a_{4} =i\omega_I( 2 \omega_A^2 +\omega_I^2 )
(k_y\hat{u}_{z0}-k_z\hat{u}_{y0})/k+i\omega_d \omega_A^2 \hat{u}_{x0}  .
\end{equation}
\end{subequations}
Solving \eqref{eq:2.4abcd} and \eqref{eq:2.5abcd} together, we obtain values for $A_{1x}, B_{1x}, C_{1x}$ and $D_{1x}$. To further solve this, we use approximate expressions for frequencies ($\varpi_{1-4}$) obtained similarly, as shown in \eqref{eq:1.8ab}. Hence, obtain the approximate analytical estimates for the Fourier-transformed velocity field ($\hat{\bm{u}}_x$)  as showcased next for both parallel and perpendicular configurations.
\subsection{Approximate analytical Energy estimates}
Consider the slow magnetostrophic wave component from  \eqref{eq:2.1a}; $\hat{{u}}_{x,s} = D_{1x} e^{{i\varpi_{4} t}}$. In the parallel configuration, it gets simplified as follows,

\begin{equation}
  \hat{u}_{x,s}\approx -\hat{u}_{x0}\dfrac{ Le^2\delta^2 k^2}{2} \exp {\left(  -\omega_d t\right) }\left(\cos\left(\frac{\omega_A^2  }{\omega_I }t\right)  -  \left(\dfrac{ 2\omega_d\omega_I}{  \omega_A^2}\right) \sin \left(\frac{\omega_A^2  }{\omega_I }t\right) \right) .
\end{equation}
The characteristic decay timescale for slow magnetostrophic wave ($\hat{u}_{x,s}$) is the magnetic dissipation time $t_\eta = \delta^2/\eta$ (see \eqref{eq:1.8ab}).\\
For $t \ll t_\eta$, 
\begin{equation}\label{eq:2.5}
    \hat{u}_{x,s}\approx -\dfrac{\hat{u}_{x0} }{2 } Le^2\delta^2 k^2.
\end{equation} A similar expression for the toroidal component of slow wave velocity for an axisymmetric problem in cylindrical polar coordinates can be seen in the work by \cite{Sreenivasan17}.

Following similar steps, for the slow magnetostrophic wave velocity calculations in the perpendicular configuration, we get,

\begin{equation}
  \hat{u}_{x,s}\approx -\hat{u}_{x0}\dfrac{ Le^2\delta^2 k_x^2k^2}{2 k_z^2} \exp {\left(  -\omega_d t\right) }\left(\cos\left(\frac{\omega_A^2  }{\omega_I }t\right)  -  \left(\dfrac{ 2\omega_d\omega_I}{  \omega_A^2}\right) \sin \left(\frac{\omega_A^2  }{\omega_I }t\right) \right) .
\end{equation}
During the initial wave phase where $t \ll t_\eta$,
\begin{equation}\label{eq:2.6}
    \hat{u}_{x,s}\approx -\dfrac{\hat{u}_{x0} }{2}  Le^2\delta^2 k^2 \times\left(\frac{k_x^2}{k_z^2}\right).
\end{equation}
By comparing the \eqref{eq:2.5} and \eqref{eq:2.6}, we observe that the slow wave velocity is scaled by a factor $k_x^2/k_z^2$ in the perpendicular configuration compared to the parallel configuration. 

Let's also consider the slow magnetostrophic wave's group velocity expression given by \cite{Davidson2001}; \cite{Finlay2010},

\begin{equation}\label{eq:2.7}
    C_g = \frac{k(\bm{B}\cdot \bm{k})^2}{2( \bm{\Omega} \cdot \bm{k}) \rho \mu}\left(\frac{\hat{\bm{k}}}{k} +\frac{2\bm{B}}{\bm{k}    \cdot\bm{B}}-\frac{\bm{\Omega}}{\bm{k}    \cdot\bm{\Omega}}\right).
\end{equation}
Substituting $\bm{B}\cdot \bm{k} = B k_z$ for the parallel configuration and $\bm{B}\cdot \bm{k} = B k_x$ for the perpendicular configuration along with  $\bm{\Omega}\cdot \bm{k} = \Omega k_z$ in \eqref{eq:2.7} we get,

\begin{equation}\label{eq:2.8a}
   \begin{aligned}
      C_{g}^{\parallel} = 2\Omega Le^2\delta^2 \left(k_z\hat{\bm{k}}+ k \hat{\bm{e}}_{\bm{z}}\right), 
    \end{aligned} \hspace{3mm}
    \begin{aligned}
        C_{g}^{\perp}= 2\Omega Le^2\delta^2 \times \frac{k_x}{k_z}\left({k_x}\hat{\bm{k}}+ 2k\hat{\bm{e}}_{\bm{x}}-k\frac{k_x}{k_z}  \hat{\bm{e}}_{\bm{z}}\right). 
    \end{aligned}
\end{equation}

We again see scaling by the factor $k_x^2/k_z^2$ in the perpendicular configuration compared to the parallel configuration (along $\bm{\hat{e}_z}$). In the perpendicular configuration, also notice that in the direction of field $|\bm{B}|\bm{\hat{e}_x}$, $C_g$ has a scaling of the order $k_x/k_z$. This shows that slow magnetostrophic wave velocities get scaled along both the rotation vector ($\bm{\Omega}$) and along the magnetic field ($\bm{B}$) in the case of $\bm{B}\perp\bm{\Omega}$. In a low Lehnert number flow ($Le \ll 1$), disturbances rapidly elongate along the rotation axis ($k_z$ decreases) and form columnar structures as propagation of weakly modified inertial waves \citep{Davidson2006}. As a consequence, we expect the $k_x/k_z$ value to be inherently large in such cases and to yield enhanced slow magnetostrophic wave velocities in the perpendicular configuration. In \eqref{eq:2.6} and for $C_g^{\perp}$ from \eqref{eq:2.8a}, $k_z \approx 0$ is the case of inertial Alfv\'{e}n waves. In the following section, with numerical simulations, we consider the inertial Alfv\'{e}n waves ($k_z \approx 0$) separately and suggest that, even when $k_z \neq  0$, the perpendicular configuration still supports a faster propagation along $\bm{B}$.


\begin{figure}[H]
\caption*{$ Le= 10^{-2},E_\eta = 10^{-4}$}
\begin{center}
\includegraphics[width = \textwidth]{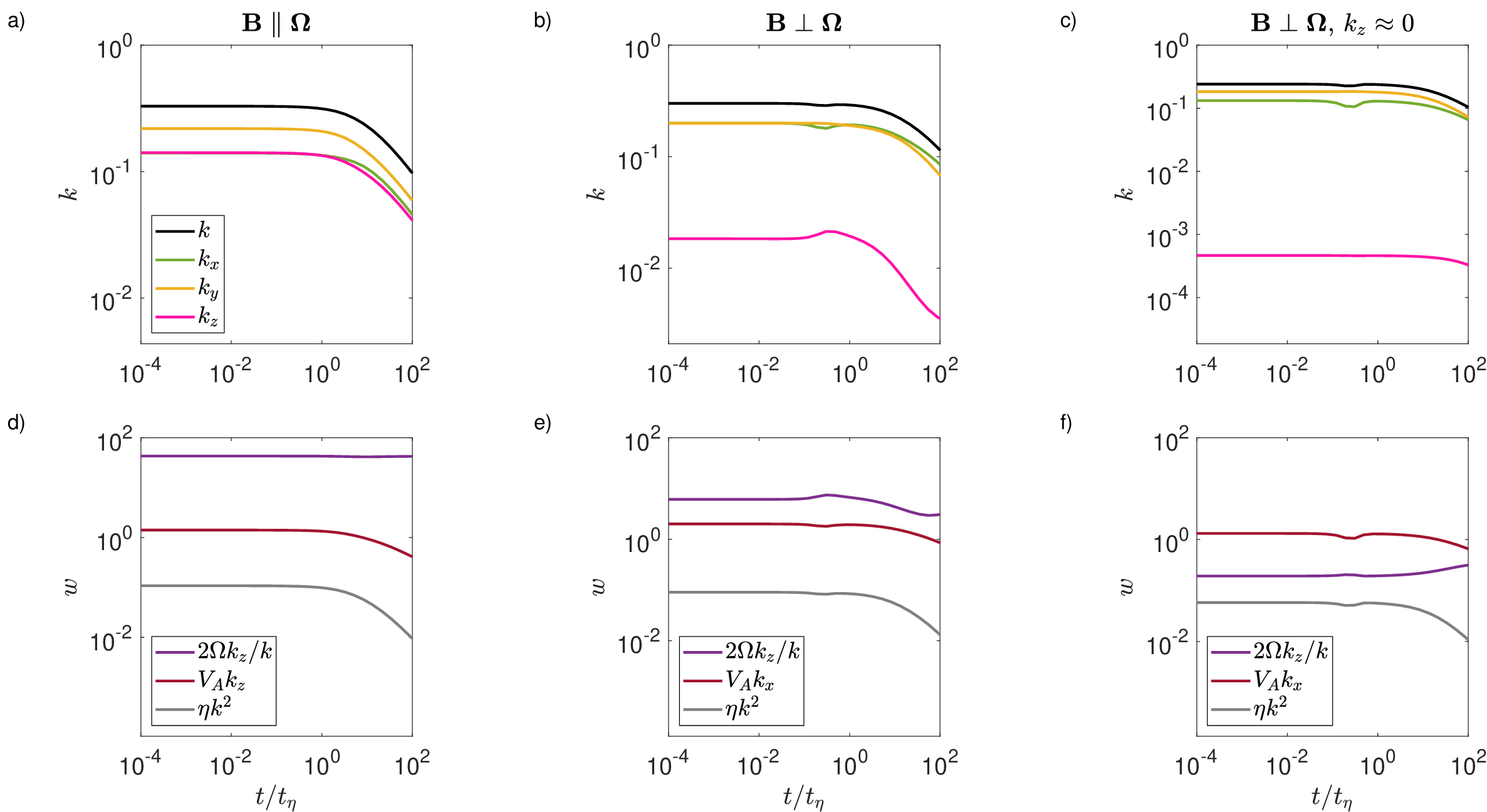}
\end{center}
\caption{Temporal evolution of wave numbers and frequencies for inertial, Alfv\'{e}n and diffusion, denoted by $2\Omega k_z/k$,  $V_A k_B$ and $\eta k^2$ respectively. With the formation of columnar structures, $k_z$ decreases (b, c) when $\bm{B} \perp \bm{\Omega}$. Appearance of inertial Alfv\'{e}n waves (f) with frequencies in the order ${\omega_I}^{\perp} \ll  {\omega_A}^{\perp}$.}
\label{fig:fig1}
\end{figure}
\begin{figure}[H]
\caption*{$E_\eta = 10^{-4}$}
\begin{center}
\includegraphics[width = 0.95\textwidth]{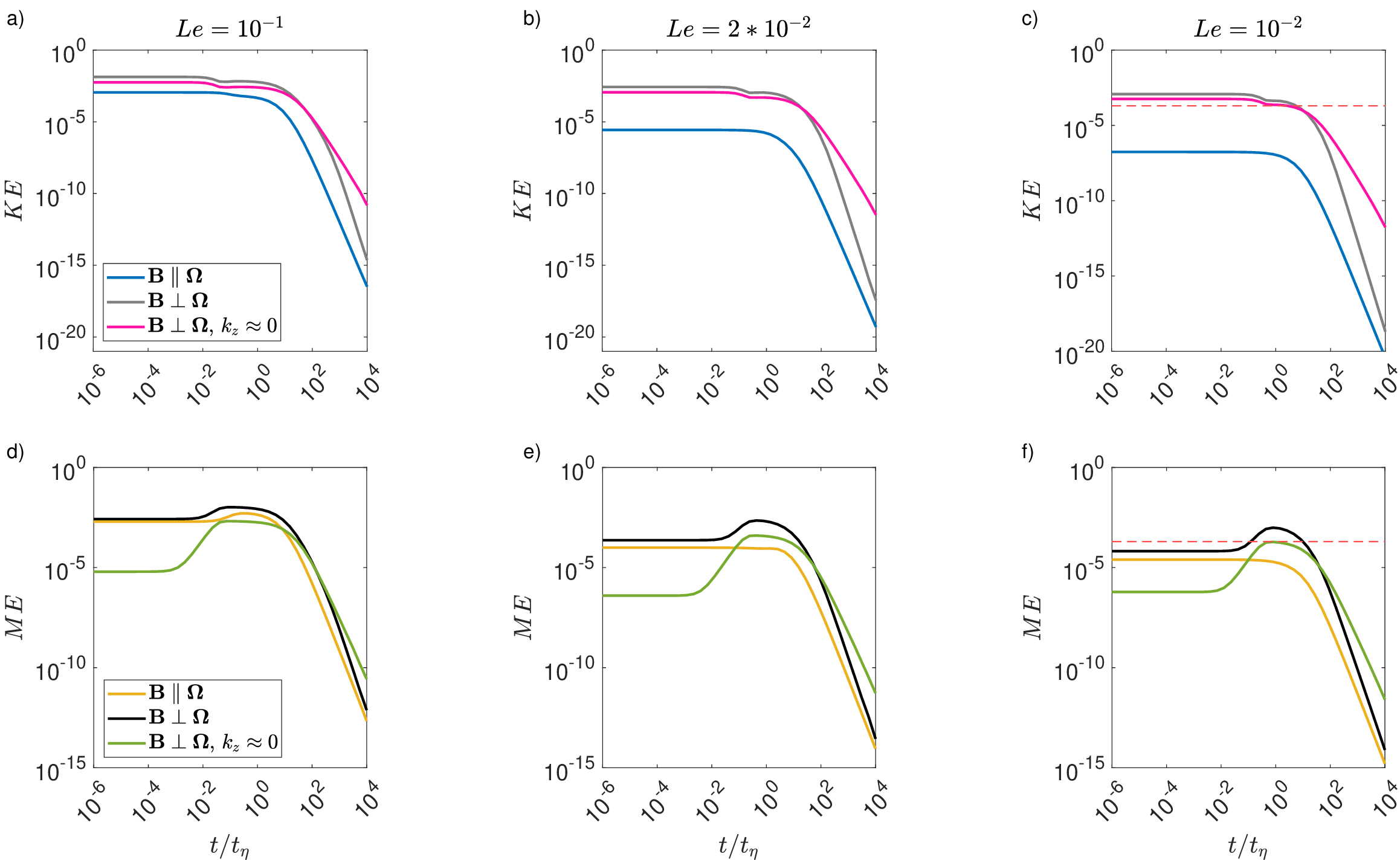}  
\end{center}
\caption{Temporal evolution of kinetic and magnetic energy (normalised with initial kinetic energy) having higher values in the case of $\bm{B} \perp \bm{\Omega}$ compared to the case $\bm{B} \parallel \bm{\Omega}$. The inertial Alfv\'{e}n waves ($k_z \approx 0$) energy values lie in between.}
\label{fig:Fig2}
\end{figure}
\section{Numerical solution to the initial value problem: velocity and energy}\label{sec:3} 
The numerical solution to $\bm{u}$, $\bm{b}$ and kinetic, magnetic energies are obtained by taking inverse Fourier transforms and applying Parseval's theorem, respectively, on  \eqref{eq:2.1a} and \eqref{eq:2.2b}. We used Simpson's $1/3^{rd}$ rule to solve these integrals, with the initial length scale, $\delta$, set to 10 m, while the magnetic diffusivity, $\eta$, was assigned a value of 1 $m^2/s$. Throughout this study we consider, $E_\eta = 10^{-4}$  and $Le= 10^{-2}-10^{-1}$. Figure \eqref{fig:fig1} shows the temporal evolution of wave numbers calculated through ratios of $L^2$ norms, 
\begin{equation}
    {k}_x = \frac{\|k_x \hat{u}_x\|}{\|\hat{u}_x\|},{k}_y = \frac{\|k_y \hat{u}_x\|}{\|\hat{u}_x\|},{k}_z = \frac{\|k_z \hat{u}_x\|}{\|\hat{u}_x\|} \hspace{2mm}\mbox{and} \hspace{2mm}{k} = \frac{\|k \hat{u}_x\|}{\|\hat{u}_x\|}.
\end{equation}
Based on this, the temporal evolution of frequencies was also obtained (figure \ref{fig:fig1}, d, e and f). Notice that in the parallel configuration, wave numbers evolve isotropically (figure \ref{fig:fig1}, a), but not in the perpendicular configuration (figure \ref{fig:fig1}, b) as suggested by \cite{Finlay2010}. This is due to the innate presence of the inertial-Alfv\'{e}n waves in the second case. We see $k_z$ drops substantially in the perpendicular case, and its effect is more pronounced when we see inertial-Alfv\'{e}n waves separately (figure \ref{fig:fig1}, c). 
For the inertial-Alfv\'{e}n waves, the condition  $\bm{k}\cdot\bm{\Omega}\approx 0$ was achieved by setting upper and lower limits in the Fourier integrals for $ k_z = \pm V_Ak_x/2\Omega\delta$, where $k_x=k/\sqrt[3]\delta$ (assuming $k_x$ does not vary in parallel and perpendicular configurations during the initial wave dominated phase). We see the discrepancy follows in the frequency plots, where inertial wave frequency $\omega_I$ drops substantially in the perpendicular configuration in comparison to the parallel case  (figure \ref{fig:fig1}, e). 
These numerical results align with the analytical calculations, which demonstrate that frequencies behave differently in both configurations, as expressed in equations \eqref{eq:1.8a} and \eqref{eq:1.8b}. In the case of inertial-Alfv\'{e}n waves  (figure \ref{fig:fig1}),  inertial frequency is lower than the Alfv\'{e}n frequency; ${\omega_I}^{\perp} \ll  {\omega_A}^{\perp}$ \citep{Bardsley16}.
\begin{figure}[H]
\caption*{$\bm{B}\parallel\bm{\Omega},\hspace{1mm} {Le=10^{-1} \& \hspace{1mm}E_\eta = 10^{-4}} $}
\begin{center}
\includegraphics[width = \textwidth]{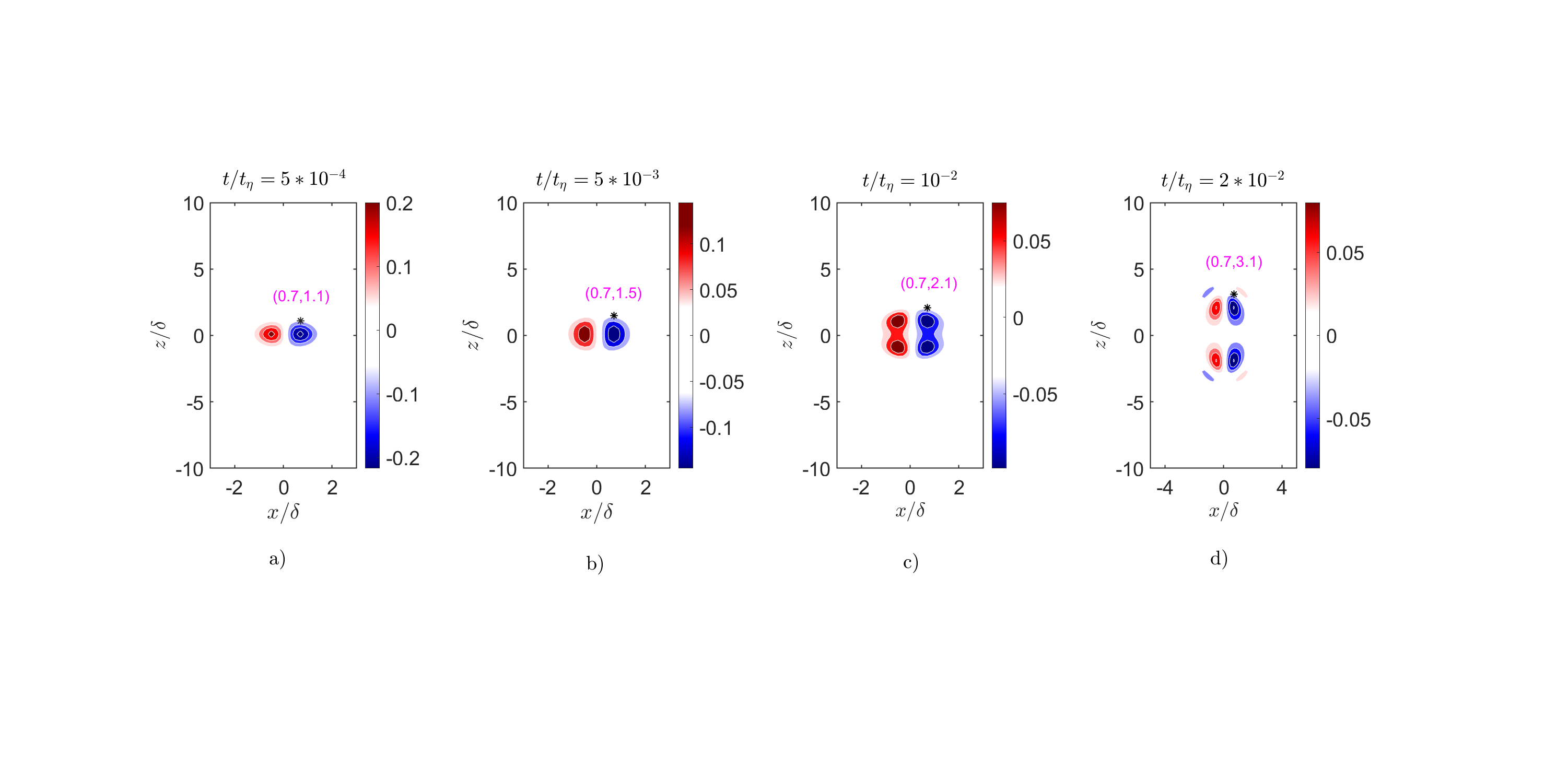}
\end{center}
\caption{Spatio-temporal evolution of $u_x$ in the $x-z$ plane for the case $\bm{B}\parallel\bm{\Omega}$, disturbances travel along the $\bm{B}$ field lines (z-axis) slowly at group velocities of magnetostrophic waves.}
\label{fig:Fig3} 
\end{figure}

Kinetic and magnetic energy for the slow wave in the case of a perpendicular configuration is substantially higher compared to the case of a parallel configuration with $Le 
\ll 1$ (Figure \ref{fig:Fig2}). Also note that inertial-Alfv\'{e}n waves ($k_z\approx0$) show a similar increment but have less energy than that calculated with the total wave vector spectrum for $k_z (-\infty,\infty)$. This suggests that the role of wave numbers $k_z \neq 0$ in enhancement of the energy can be interpreted from \eqref{eq:2.6} and \eqref{eq:2.8a}. Following the work by \cite{Davidson2017}, there is an existence of $k_z$ values leading to intermediate magnetic-Coriolis waves where ${\omega_I}^{\perp} \approx  {\omega_A}^{\perp}$ though $Le \ll 1$, which are responsible for the increase in energies. Their study also proposes that the wave associated with $k_z \neq 0$ should decay faster compared to inertial-Alfv\'{e}n waves ($k_z \approx 0$). This gets demonstrated in this study as after diffusion time $t/t_\eta \approx1$, slow waves' energies with $k_z (-\infty,\infty)$ decay fast compared to inertial-Alfv\'{e}n waves' energy ($k_z \approx 0$). We see equipartition of kinetic and magnetic energy for the inertial-Alfv\'{e}n waves (dashed red lines in figure \ref{fig:Fig2}, c,f ). It is also interesting to note that decay starts on the same time scales ($t=t_\eta$) in both parallel and perpendicular configuration, as expected from \eqref{eq:1.8a} and \eqref{eq:1.8b}.

\begin{figure}[H]
\caption*{$\bm{B}\perp\bm{\Omega},\hspace{1mm} {Le=10^{-1} \& \hspace{1mm}E_\eta = 10^{-4}} $}
\begin{center}
\includegraphics[width = 0.6\textwidth,height = 0.7\textwidth]{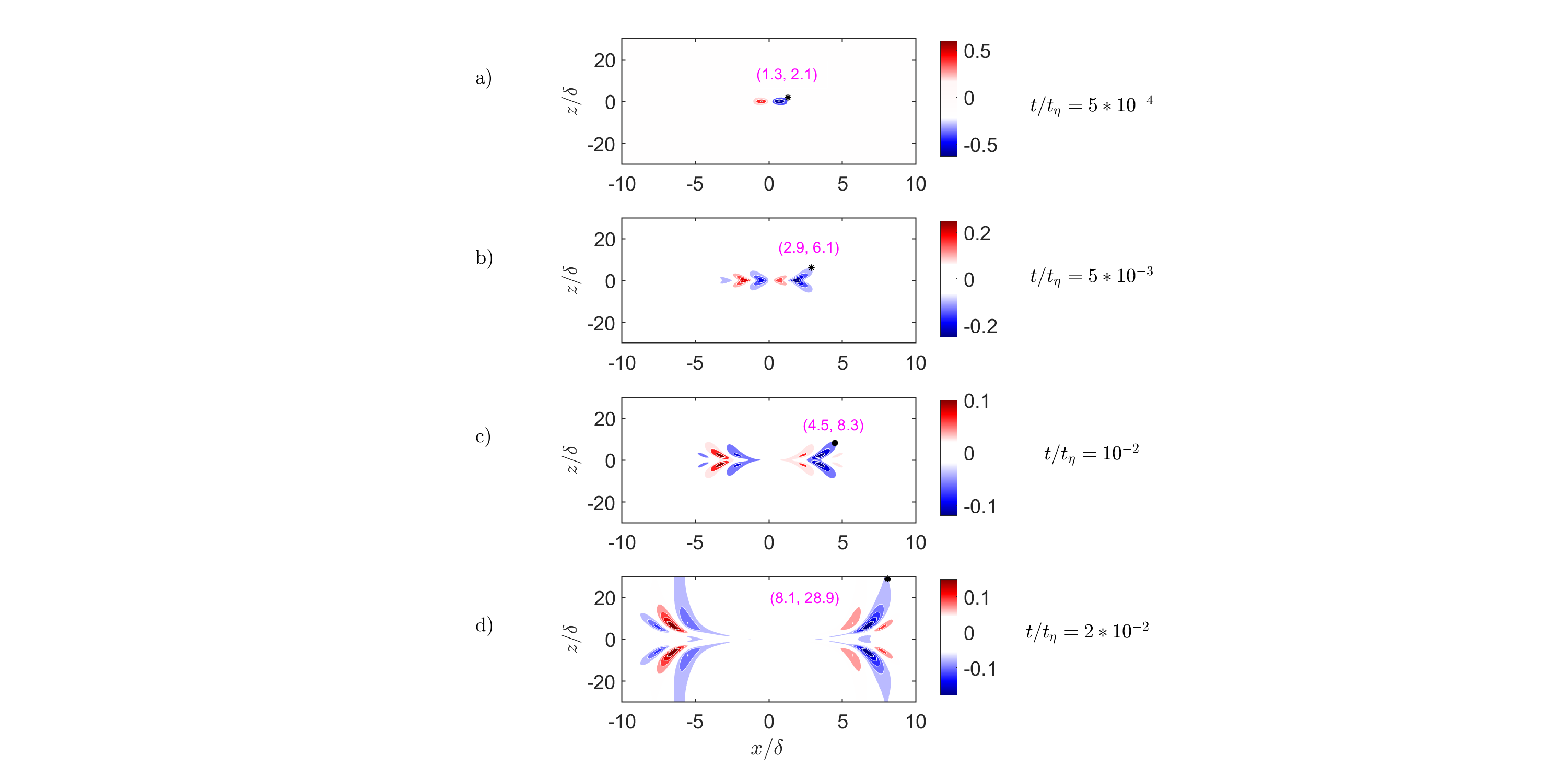}
\end{center}
\caption{Spatio-temporal evolution of $u_x$ in the $x-z$ plane for the case $\bm{B}\perp \bm{\Omega}$, disturbances travel along the $\bm{B}$ field lines (x-axis) fast with velocities comparable to the Alfv\'{e}n velocity. Simultaneous fast evolution along the rotation axis is attributed to the formation of inertial Alfv\'{e}n waves through wave-numbers $k_z\approx 0$.}
\label{fig:Fig4}
\end{figure}

In the parallel configuration, slow wave velocities evolve along the z-axis with velocities much less than the Alfv\'{e}n velocity $V_A$ (figure \eqref{fig:Fig3}). Compared to this, at the same stage of temporal evolution, the perpendicular configuration exhibits much faster evolution along the field lines (x-axis) and also along the rotation axis (Figure \ref{fig:Fig4}). Velocity of evolution along the x-axis is very close to  Alfv\'{e}n velocity in this case $V_A = 100$. This suggests the combined influence of both inertial-Alfv\'{e}n waves and intermediate magnetic-Coriolis waves in the enhancement of velocities along the field lines. Rapid evolution along the rotation axis in the perpendicular configuration is due to the formation of inertial-Alfv\'{e}n waves, which are absent in the parallel configuration.


\section{Conclusion}\label{sec:4}

Analytical derivations for magnetostrophic waves' frequency expression in \eqref{eq:1.8ab} show that there is a difference in the wave's phase velocity depending on the orientation of the magnetic field $\bm{B}$ and rotation vector $\bm{\Omega}$  \citep{Finlay2010};  \citep{Hori2022}. Corresponding numerical results plotted in Figure \ref{fig:fig1} additionally show that in the perpendicular configuration, we see that inertial and Alfv\'{e}n frequencies are comparable ($\omega_I \approx\omega_A$) relatively to those in the case of parallel configuration ($\omega_I \gg\omega_A$). This supports the existence of intermediate magnetic-Coriolis waves and also inertial Alfv\'{e}n waves in the perpendicular configuration \citep{Davidson2017}.

Comparing the imaginary part of the equations \eqref{eq:1.8a} and \eqref{eq:1.8b}, we observe that the timescales for decay of these waves remain the same in both parallel and perpendicular configurations. This can also be observed in Figure \ref{fig:Fig2}, which shows the numerical evolution of kinetic and magnetic energy, and these start to decay at the same instance ($t=t_\eta$) in both configurations. Interestingly, inertial Alfv\'{e}n waves happen to have decay time scales half that of the magnetostrophic wave \eqref{eq:1.7}, and start to decay earlier than magnetostrophic waves.

Formation of columnar structures perpendicular to the magnetic field gives enhanced magnetostrophic wave velocities with scaling $k_x^2/k_z^2$ in perpendicular configuration \eqref{eq:2.6} compared to the case when these columnar structures form along the magnetic field in parallel configuration  \eqref{eq:2.5}. Comparison of group velocities in the two cases \eqref{eq:2.8a} further shows that this enhancement is by a scale factor of $k_x^2/k_z^2$ along the rotation vector (z-axis) and by a reduced factor of $k_x/k_z$ along the magnetic field (x-axis). Velocity counter plotted based on numerical calculations in Figure $\ref{fig:Fig3}$ (parallel configuration) and Figure \ref{fig:Fig4} (perpendicular configuration) supports these analytical estimates.
 
 Numerical results for energy estimates show that in the initial wave-dominated phase, the magnetostrophic wave components exhibit a remarkable increase in energy values (see Figure \ref{fig:Fig2}). Specifically for an Earth-like Lehnert number of $Le = 10^{-2}$ and magnetic Ekmann number $E_\eta = 10^{-4}$, the magnetostrophic wave kinetic energy shows an increase of $O(10^4)$, while the magnetostrophic wave magnetic energy increases by $O(10^2)$. The slow magnetostrophic wave magnetic field intensity in the perpendicular configuration is approximately six times greater than that in the parallel configuration. This demonstrates that strong wave motion occurs along the azimuthal direction due to slow magnetostrophic waves with $\bm{B}$ orthogonal to $\bm{\Omega}$, which also suggests that a low-intensity magnetic field inside the core could sustain the propagation of these waves.

\backsection[Acknowledgements]{The authors thank Prof. Binod Sreenivasan (CEaS, IISc, Bangalore), Prof. Avishek Ranjan (Dept. Mechanical Engineering, IIT, Bombay) and Prof. Jai Sukhatme (CAOS, IISC, Bangalore) for their support and helpful comments.}

\appendix

\section{}\label{appA}
\subsection{Approximate frequency for the slow magnetostrophic wave}\label{appA1}
Taylor series expansion for terms in the square root of \eqref{eq:1.8},
\begin{equation*}
    \sqrt{1+\left(\dfrac{2\omega_A}{\omega_I}\right)^2 \mp i\left( \dfrac{2\omega_d}{\omega_I}\right)}\approx 1+\frac{2\omega_A^2}{\omega_I^2} \mp i \dfrac{\omega_d}{\omega_I}\left(1-\frac{2\omega_A^2}{\omega_I^2}\right),
\end{equation*}
substituting back in \eqref{eq:1.8},
\begin{equation*}
    \varpi_{3,4} \approx \pm\dfrac {\omega_I}{2} +i \dfrac{ \omega_d}{2} \mp \dfrac {\omega_I}{2} \left[1+\frac{2\omega_A^2}{\omega_I^2} \mp i \dfrac{\omega_d}{\omega_I}\left(1-\frac{2\omega_A^2}{\omega_I^2}\right)\right] ,
\end{equation*}
by neglecting the last term with relatively small magnitude,
\begin{equation}\label{eq:A1}
  \varpi_{3,4} \approx  \mp\dfrac{ \omega_A^2}{\omega_I}+i\omega_d.
\end{equation}

\bibliographystyle{jfm}

\end{document}